\documentclass[a4paper,byrevtex,nofootinbib,showkeys,dvips]{revtex4}
\usepackage{dcolumn}
\usepackage{amsmath}
\usepackage{amssymb}
\usepackage{bm}

\usepackage{graphicx}

\begin{document}

%\draft
\title{The envelope theory as a pedagogical tool}

\author{Claude \surname{Semay}}
\email[E-mail: ]{claude.semay@umons.ac.be}
\thanks{ORCiD: 0000-0001-6841-9850}

\author{Maud \surname{Balcaen}}
\email[E-mail: ]{maud.balcaen@student.umons.ac.be}

\affiliation{Service de Physique Nucl\'{e}aire et Subnucl\'{e}aire,
Universit\'{e} de Mons,
UMONS Research Institute for Complex Systems,
Place du Parc 20, 7000 Mons, Belgium}

\date{\today}
\begin{abstract}
The envelope theory is a reliable and easy to implement method to solve time independent Schr\"odinger-like equations (eigenvalues and eigenvectors). It is mainly useful to solve many-body systems since the computational cost is independent from the number of particles.  Due to its simplicity, this method can also be used as a pedagogical tool. This is shown here for the soft-Coulomb potential $-k/\sqrt{x^2+d^2}$ in one dimension, characterised by a bias distance $d$. Such interaction is used for the study of excitons, electron-hole bound pairs where the two charges are kept separated in two different one-dimensional regions (quantum wires). 
\end{abstract}


\maketitle

\section{Introduction}
\label{sec:intro}

Finding the eigenvalues and eigenvectors of Hamiltonians is a fundamental problem in quantum mechanics. Various methods can be used as the variational one, the WKB approximation or the perturbation theory, for instance \cite{flug99,griff18}. All these methods rely on the construct of good approximations for the wavefunctions. The principle is different for the envelope theory (ET) in which the Hamiltonian under study $H$ is approximated by an auxiliary Hamiltonian $\tilde H$ whose solutions are exactly known \cite{hall80,hall83,sema13,chev21}. The approximate eigenvalues of $H$ are obtained by an extremisation procedure applied to the eigenvalues of the auxiliary Hamiltonian. A practical choice for $\tilde H$ is a harmonic oscillator Hamiltonian. If the accuracy is not great, the method is reliable and analytical upper or lower bounds can be obtained in favourable situations. The ET is mainly useful for many-body systems since the computational cost is independent from the number of particles, but it can also be used as a pedagogical tool as it is very easy to implement and allows the computation of the whole spectrum. 

In this paper, the ET is applied to the study of the soft-Coulomb potential $-k/\sqrt{x^2+d^2}$ in one dimension, characterised by a bias distance $d$. This interaction is chosen for three reasons: (i) It is characteristic of excitons, electron-hole bound pairs where the two charges are kept separated in two different one-dimensional regions. These systems, named quantum wires, appear in solid state physics \cite{schr15} and in various biological processes \cite{mara18}; (ii) It is bounded from below and from above by two potentials for which known exact solutions can be used as control lower and upper bounds ; (iii) It is studied in a recent paper by the variational method \cite{gras17} which allows interesting comparisons. 

The recipe to use the ET for a two-body system in one dimension is presented in Sect.~\ref{sec:ET}. The ET is applied to the one-dimensional soft-Coulomb potential $-k/\sqrt{x^2+d^2}$ in Sect.~\ref{sec:exciton}, where eigenvalues and eigenvectors are computed and compared with results coming from accurate numerical calculations and from a variational method. Concluding remarks are given in Sect.~\ref{sec:conclu}.

\section{The envelope theory}
\label{sec:ET}

In this section, matters from several papers \cite{sema13,sema19,cimi21} are collected and summarised to describe how to compute the approximate ET solutions for the generic two-body Hamiltonian given by
\begin{equation}
\label{TpV}
H=T(\hat p)+ V(x),
\end{equation}
where $x$ is the relative position, $\hat p=-i\hbar\, d/dx$ the momentum conjugated with $x$, $T$ the kinetic term and $V$ the potential. In the following, it is necessary to make the difference between the momentum as an operator, denoted by $\hat p$, and the momentum as a simple variable, denoted by $p$. The domain of $x$ can be $\mathbb{R}_+$ or $\mathbb{R}$. In this last case, $V$ is an even function of $x$ for physical and practical reasons. $T$ is always expected to be an even function of $p$ \cite{sema16}. The approximate energy $E$ for the $n$th level is given by an extremisation procedure. But it can be shown that it is equivalent to solve the following system for each value of the quantum number $n$ \cite{sema13}
\begin{align}
\label{ET1}
E&=T(p_0)+ V(x_0), \\
\label{ET2}
x_0 p_0 &= Q_n \hbar, \\
\label{ET3}
p_0 T'(p_0) &= x_0 V'(x_0),
\end{align}
where $X'(z)=dX/dz$, $x_0 >0$, $p_0 >0$ and $Q_n$ is defined by
\begin{align}
\label{Q1}
Q_n&=n+\frac{1}{2} \quad \textrm{for} \quad x \in \mathbb{R} , \\
\label{Q2}
Q_n&=n_\textrm{o}+\frac{1}{2} \quad \textrm{for} \quad x \in \mathbb{R}_+ ,
\end{align}
where $n=0,1,2,3,\ldots$ and $n_\textrm{o}=2 n+1 = 1,3,5,7,\ldots$ The restriction to odd numbers $n_\textrm{o}$ insures that the wavefunction vanishes at $x=0$. The structure of $Q_n$ comes from the use of the harmonic oscillator Hamiltonian as the auxiliary Hamiltonian. Consequently, the corresponding approximate ET wavefunction is an harmonic oscillator one. For $x \in \mathbb{R}$, it is written \cite{griff18}
\begin{equation}
\label{wfET}
\langle x | n \rangle = \psi_n(x)=\left( \frac{\lambda^2}{\pi} \right)^{1/4} \frac{1}{\sqrt{2^n n!}}H_n(\lambda x)\exp\left( -\lambda^2 x^2/2 \right),
\end{equation}
where $H_n$ is an Hermite polynomial and $\lambda=\sqrt{Q_n}/x_0$. For $x \in \mathbb{R}_+$, $\psi_{n_\textrm{o}}(x)$ must be multiplied by $\sqrt{2}$ to keep the correct normalisation. It can be shown that 
\begin{equation}
\label{x0p0}
\langle n |x^2| n \rangle = x_0^2 \quad \textrm{and} \quad \langle n |\hat p^2| n \rangle = p_0^2.
\end{equation}
So, $p_0$ can be interpreted as the mean momentum of each particle and $x_0$ as the mean distance between the two particles. These physical quantities appear in three equations giving the definition of the energy (\ref{ET1}), the rule for the quantisation (\ref{ET2}) and the equation of motion (\ref{ET3}) as a transcendental equation. If this system has a nice semi-classical interpretation \cite{sema13}, the ET is a full quantum method giving eigenvectors and eigenvalues. In natural units, $\hbar = 1$ in (\ref{ET2}).

The structure of $T$ is usually simpler that the one of $V$. That is the reason why it is generally more convenient to work in the configuration space. But the symmetry between the momentum and the relative distance in the system (\ref{ET1})-(\ref{ET3}) indicates that computations can be performed indifferently in the configuration space (with $\hat p=-i\hbar\, d/dx$) or in the momentum space (with $\hat x=i\hbar\, d/dp$). 

In the ET, for each level $n$, the kinetic part $T(p)$ and the potential part $V(x)$ are respectively ``enveloped" by the following tangent quadratic functions (see the example in Sect.~\ref{sec:ETapp})
\begin{align}
\label{Ttilde}
\tilde T(p)&=T(p_0)+ \frac{T'(p_0)}{2 p_0}\left( p^2 - p_0^2 \right), \\
\label{Vtilde}
\tilde V(x)&=V(x_0)+ \frac{V'(x_0)}{2 x_0}\left( x^2 - x_0^2 \right).
\end{align}
This is the origin of the name of the method \cite{hall80,hall83}. The enveloping functions are such that $\tilde T(\hat p)+\tilde V(x)=\tilde H$ which is the auxiliary Hamiltonian for the $n$th level ($p_0$ and $x_0$ depend on $n$ by the set of equations~(\ref{ET1})-(\ref{ET3})). It can be checked from (\ref{x0p0})-(\ref{Vtilde}) that $E=\langle n |\tilde H| n \rangle$. If $T(p)$ ($V(x)$) is quadratic, $\tilde T(p)=T(p)$ ($\tilde V(x)=V(x)$). By defining the two functions $b_T(p^2)=T(p)$ and $b_V(x^2)=V(x)$, it can be shown that $E$ is an upper (lower) bound of the genuine eigenvalue if $b_T''$ and $b_V''$ are both concave (convex) functions. If the second derivative is vanishing for one of these functions, the variational character is solely ruled by the convexity of the other one. It is easy to check that the situation  $b_T''=b_V''=0$ corresponds to the harmonic oscillator Hamiltonian for which the system (\ref{ET1})-(\ref{ET3}) gives the exact solution. In the other cases, the variational character of the solution cannot be guaranteed. Let us note that $E$ has an analytical form for a large variety of Hamiltonians for arbitrary values of the dimension \cite{sema19, sema15a}.

\section{The exciton Hamiltonian}
\label{sec:exciton}

The Hamiltonian $H_\textrm{eh}$ for the electron-hole pair is given by
\begin{equation}
\label{HExc}
H_\textrm{eh}=\frac{\hat p^2}{2 m} -\frac{k}{\sqrt{r^2+d^2}},
\end{equation}
where $r$ ($\in \mathbb{R}$) is the relative position between the electron and the hole, $\hat p$ the conjugate momentum, $m$ the reduced mass of the pair and $k=e^2/(4\pi\epsilon_0\epsilon_r)$, with $e$ the elementary electric charge, $\epsilon_0$ the dielectric permittivity in vacuo and $\epsilon_r$ the relative dielectric permittivity of the material. 

\subsection{Dimensionless analysis}
\label{sec:dim}

It is always interesting to work with dimensionless variables in order to make apparent what are the relevant physical quantities and to simplify the formulation of the problem. By using the effective Bohr radius $a_\textrm{B}=4\pi\epsilon_0\epsilon_r\hbar^2/(m e^2)$ and the effective Rydberg constant $\textrm{R}_\textrm{y}=m e^4/(32\pi^2\epsilon_0^2\epsilon_r^2\hbar^2)=k/(2 a_\textrm{B})$, a dimensionless Hamiltonian $H=H_\textrm{eh}/(2 \textrm{R}_\textrm{y})$ can be defined in terms of a dimensionless bias parameter $D=d/a_\textrm{B}$ and a dimensionless position $x=r/a_\textrm{B}$
\begin{equation}
\label{Hrename}
H=-\frac{1}{2} \frac{d^2}{dx^2}-\frac{1}{\sqrt{x^2+D^2}},
\end{equation}
whose dimensionless eigenvalues will be designed by $E$. Only dimensionless quantities will be used in the following. To recover physical quantities, $x$ must be multiplied by $a_\textrm{B}$ and $E$ by $2 \textrm{R}_\textrm{y}$. Both $a_\textrm{B}$ and $\textrm{R}_\textrm{y}$ are pertinent units for contexts in which excitons appear. $D$ is around unity for typical semiconductors.

\subsection{Bounding potentials}
\label{sec:BV}

\begin{figure}[htb]
\includegraphics[width=7cm]{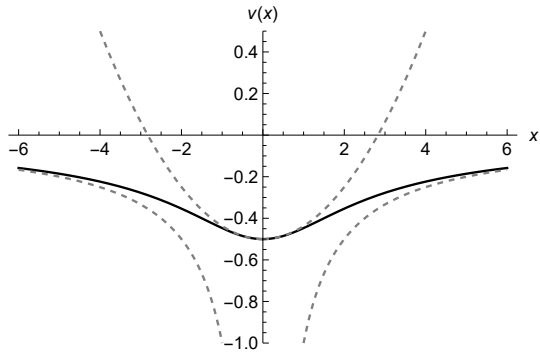}
\caption{The soft-Coulomb potential for $D=2$ (solid black) with the Coulomb potential (dashed gray) and the quadratic potential with $D=2$ (dashed gray). \label{fig:VB}}
\end{figure}

The soft-Coulomb potential $-1/\sqrt{x^2+D^2}$ is bounded from below by the Coulomb potential $-1/|x|$ for $x\in\mathbb{R}$ (see Fig.~\ref{fig:VB}). So, according to the comparison theorem \cite{sema11}, the eigenvalues of 
\begin{equation}
\label{HCoul}
H_\textrm{C}=-\frac{1}{2} \frac{d^2}{dx^2}-\frac{1}{|x|}
\end{equation}
are lower bounds of the eigenvalues of $H$. The spectrum of $H_\textrm{C}$ is analytical \cite{jara09}
\begin{equation}
\label{ECoul}
E_\textrm{C}=-\frac{1}{2(n+1)^2}, 
\end{equation}
with $n=0,1,2,3,\ldots$ associated with wavefunctions vanishing at $x=0$ and with $2n$ nodes at finite $x$ values on $\mathbb{R}$. To correspond to our numbering scheme (\ref{Q1})-(\ref{Q2}), (\ref{ECoul}) must be rewritten 
\begin{equation}
\label{ECoul2}
E_\textrm{C}=-\frac{2}{(n_\textrm{o}+1)^2}.
\end{equation}
For large values of $n_\textrm{o}$, the eigenfunctions of $H$ and $H_\textrm{C}$ are characterised by a large extension, that is to say a significant part in regions where both potentials are very similar. So it is expected that $E$ and $E_\textrm{C}$ will be very close in these situations.

An easy way to solve approximately a Schr\"odinger equation around an equilibrium point is to use a harmonic approximation of the potential. Expanding the soft-Coulomb potential in power of $x$ around 0, $H$ reduces to the harmonic oscillator Hamiltonian
\begin{equation}
\label{HHO}
H_\textrm{HO}=-\frac{1}{2} \frac{d^2}{dx^2}+ \frac{x^2}{2 D^3}-\frac{1}{D}
\end{equation}
at the lowest order. The corresponding eigenvalues are given by
\begin{equation}
\label{EHO}
E_\textrm{HO}=\left( n+\frac{1}{2} \right)\frac{1}{D^{3/2}} -\frac{1}{D}.
\end{equation}
It is easy to show that the quadratic potential in $H_\textrm{HO}$ is always larger that the one in $H$ (see Fig.~\ref{fig:VB}). So, according to the comparison theorem, $E_\textrm{HO}$ are upper bounds of $E$. Moreover, both potentials are very similar for small values of $x$. It is then expected that $E$ and $E_\textrm{HO}$ will be very close for values of $n$ close to 0, corresponding to eigenfunctions characterised by a small extension.

\subsection{ET approximations}
\label{sec:ETapp}

The ET solutions for $H$ are computed by solving first (\ref{ET2})-(\ref{ET3}) for a nonrelativistic kinematics and the soft-Coulomb potential. The equation for $x_0$ is then
\begin{equation}
\label{x0SC}
x_0^4 = Q_n^2 \left(x_0^2 + D^2 \right)^{3/2}.
\end{equation}
Analytical solution for $x_0$ can be written in terms of the solutions of a quartic equation, but the expression is so complicated that it is not usable in practice. Graphically, it is easy to check that (\ref{x0SC}) has only one positive solution for $x_0^2$. It is then very easy to compute it by a numerical procedure or a software like Mathematica\textsuperscript{\textregistered}. The ET approximations $E_\textrm{ET}$ for the energies are upper bounds since $b_T''=0$ and $b_V(z)=-(z+D^2)^{-1/2}$ is a concave function. Finally, the ET upper bounds are given by
\begin{equation}
\label{EET}
E_\textrm{ET}=\frac{Q_n^2}{2 x_0^2}-\frac{1}{\sqrt{x_0^2+D^2}} \quad \textrm{with} \quad x_0^4 = Q_n^2 \left(x_0^2 + D^2 \right)^{3/2}.
\end{equation}

As presented in Sect.~\ref{sec:BV}, the solutions for $H$ must be close to the solutions for $H_\textrm{C}$ and $H_\textrm{HO}$ under certain conditions. Let us check this from (\ref{EET}). Provided $D$ is not too large, it is expected from (\ref{x0SC}) that $x_0^4 \approx Q_n^2 x_0^3$, that is to say $x_0 \approx Q_n^2$, when $n \gg 1$. In this case, the parameter $D$ disappears in the dominant term and (\ref{EET}) reduces to 
\begin{equation}
\label{EETng}
E_\textrm{ET} \approx -\frac{2}{(2n+1)^2}.
\end{equation}
This last formula can be compared with $E_\textrm{C}$ for odd values of $n$: (\ref{EETng}) is clearly an upper bound of (\ref{ECoul2}).

If $n$ is a small integer and $D$ is not too small, an approximate solution for (\ref{x0SC}) is $x_0^4 \approx Q_n^2 D^3$, that is to say $x_0^2 \approx Q_n D^{3/2}$. In this case, at the lowest order in $n$, (\ref{EET}) reduces to 
\begin{equation}
\label{EETns}
E_\textrm{ET} \approx \left( n+\frac{1}{2} \right)\frac{1}{D^{3/2}} -\frac{1}{D},
\end{equation}
which is identical to $E_\textrm{HO}$. This is not surprising since the ET gives the exact solutions for a harmonic oscillator Hamiltonian.

\begin{figure}[ht]
\includegraphics[width=7cm]{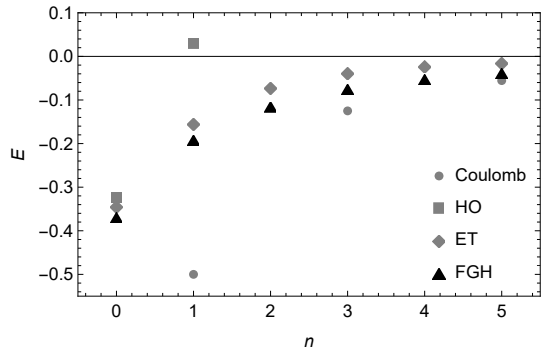}
\caption{Lowest energies for $H$ with $D=2$: accurate values obtained with the FGH method (black triangle) and upper bounds from the ET (gray diamond). The exact energies for $H_\textrm{C}$ (gray circle) and $H_\textrm{HO}$ with $D=2$ (gray square) are also indicated. \label{fig:En}}
\end{figure}

\begin{figure}[ht]
\includegraphics[width=7cm]{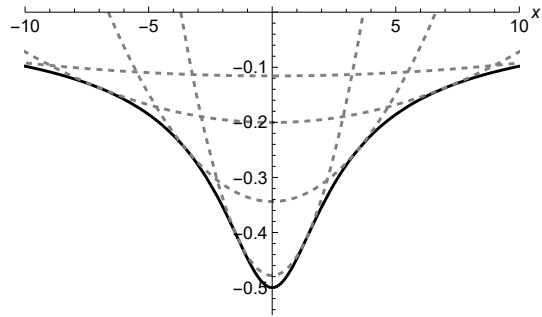}
\caption{Four first envelopes $\tilde V$ (dashed gray) for $n=\{0,1,2,3\}$ of the soft-Coulomb potential (solid black) for $D=2$. \label{fig:Env}}
\end{figure}

\begin{figure}[htb]
\includegraphics[width=15cm]{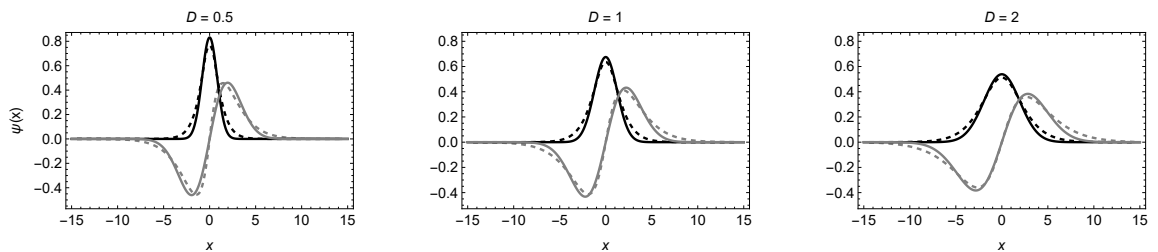}
\caption{Ground (black) and first excited (gray) states of $H$ for three values of $D$: accurate wavefunctions computed with the FGH method (dashed), and approximate wavefunctions computed with the ET (solid). \label{fig:WF}}
\end{figure}

Accurate numerical eigenvalues and eigenvectors of $H$ are computed with the Fourier Grid Hamiltonian (FGH) method \cite{mars89,sema00} which is very powerful and very easy to use for one-dimensional time independent Schr\"odinger-like equations. In Fig.~\ref{fig:En}, theses energies are compared with $E_\textrm{ET}$ and the other bounds $E_\textrm{HO}$ and $E_\textrm{C}$, for $D=2$ and the lowest values of $n$. The ET approximations are good for all values of $n$. The lower bound $E_\textrm{C}$ which is only defined for odd numbers $n$ improves with increasing $n$, as expected. On the contrary, the upper bound $E_\textrm{HO}$ degrades rapidly with increasing $n$, as expected also. It is worth drawing attention that the upper bounds $E_\textrm{HO}$ are computed with a unique quadratic potential in (\ref{HHO}) (see Fig.~\ref{fig:VB}). The upper bounds $E_\textrm{ET}$ are computed with a different envelope potential $\tilde V(x)$ for each value of $n$ (see Fig.~\ref{fig:Env}). Wavefunctions for the ground state and the first excited state, computed with the FGH method and the ET, are compared in Fig.~\ref{fig:WF} for three values of $D$. The agreement is good but the ET approximations decrease faster with $x$. This is due to the harmonic oscillator nature of the approximate wavefunctions which decrease as Gaussian functions while the exact solutions decrease as exponential functions. 

\subsection{Comparison with a variational method}
\label{sec:comp}

\begin{figure}[htb]
\includegraphics[width=7cm]{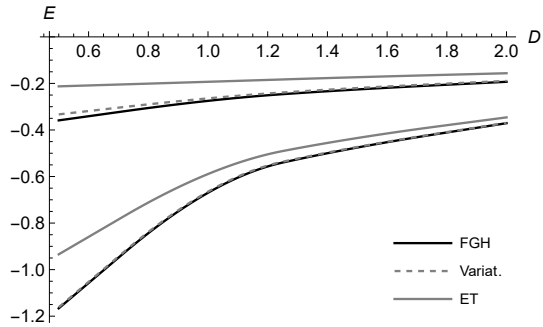}
%\quad \includegraphics[width=7cm]{FigD2.eps}
\caption{Energies of the ground (3 curves below) and first excited (3 curves above) states of $H$ as un function of $D$: accurate values obtained with the FGH method (solid black), upper bounds from the variational method in \cite{gras17} (dashed gray), upper bounds from the ET (solid gray). 
%[Right] Relative errors $\Delta$ for the upper bounds of $H$ from the ET for the ground (black) and first excited (gray) states as un function of $D$. 
\label{fig:D}}
\end{figure}

The ground and first excited states of $H$ have been computed in a recent paper by a variational method using the ground and first excited states of the harmonic oscillator Hamiltonian as trial states \cite{gras17}. The results from this paper are compared with the ET results in Fig.~\ref{fig:D} as a function of $D$. Let us mention that the dimensionless numbers obtained with the formulas in \cite{gras17} must be divided by 2 to be compared with our results, because the energies are given in unit of $\textrm{R}_\textrm{y}$, while they are given in unit of $2 \textrm{R}_\textrm{y}$ in this paper (see~(\ref{Hrename})). It is clear that the variational upper bounds are better that the upper bounds given by the ET.  Nevertheless, the ET can give the whole spectra with the same computational cost. This is not the case in \cite{gras17}, where different integrations must be performed to compute ground and first excited states. Moreover, the variational principle states that the expectation value of a Hamiltonian for an arbitrary trial state always gives an upper bound of the ground state. But an upper bound of an excited state can be reliably computed with a trial state if it is orthogonal to \textbf{all exact states} below this excited state \cite{griff18}. So, for a one dimensional system, an upper bound of the ground state can be computed with an even trial wave function and an upper bound of the first excited state with an odd trial wave function. The computation of the other excited states is much more complicated and requires the expansion of trial states in an (orthonormal) basis \cite{macd33}.

\section{conclusions}
\label{sec:conclu}

The envelope theory (ET) is useful to treat problems with a large number of particles when a great accuracy is not required \cite{buis12,sema21,buis22}. It can also be used to produce test calculations for numerical methods more accurate but more difficult to implement \cite{horn14,timo17}. Nevertheless, it can also be used in a pedagogical context to study solutions of time independent Schr\"odinger-like equations, for the following reasons:
\begin{itemize}
  \item The ET is very simple to implement since the solutions are obtained by solving a transcendental equation, and the generalisation for many-body systems in several dimensions is trivial \cite{sema13}.
  \item The eigenvalues and eigenvectors can be computed for the whole spectrum with the same computational cost, and limiting cases studied. 
  \item In the most favourable situations, analytical upper or lower bounds can be computed. 
\end{itemize}
The calculations performed above for the one-dimensional soft-Coulomb potential can be repeated for a lot of different interactions and/or for a semirelativistic kinematics. An interesting case is the Hulth\'en potential on $\mathbb{R}_+$, $-k/\left(\exp{(a x)}-1\right)$, which is bounded from above by $-k \exp{(-a x)}$ and from below by $-k /(a x)$. Analytical solutions are known for these three interactions in a one-dimensional Schr\"odinger equation \cite{flug99}.  

Let us mention that the results of the ET in dimensions greater than one can be improved by combining it with the dominantly orbital state method \cite{chev21,sema15b}. The idea is to modify the equivalent of the quantum number $Q_n$ appearing in these contexts. Such improvement could be possible in one dimension also by replacing $n+1/2$ by $n+\gamma(n)$ in $Q_n$, in the same spirit as the modification of the WKB method proposed in \cite{delv21}. The function $\gamma(n)$ could be fitted on accurate numerical results to produce better upper bounds. Similar calculations are already performed in \cite{sing85}.

%\begin{acknowledgments}
%\end{acknowledgments} 


\begin{thebibliography}{aa}

\bibitem{flug99} S. Fl\"ugge, \emph{Practical Quantum Mechanics} (Springer, New York, 1999)
\bibitem{griff18} D.J. Griffiths, D.F. Schroeter, \emph{Introduction to Quantum Mechanics} (Cambridge University Press, Cambridge, 2018)
\bibitem{hall80} R.L. Hall, Energy trajectories for the $N$-boson problem by the method of potential envelopes, Phys. Rev. D \textbf{22}, 2062 (1980)
\bibitem{hall83} R.L. Hall, A geometrical theory of energy trajectories in quantum mechanics, J. Math. Phys. \textbf{24}, 324 (1983) 
\bibitem{sema13} C. Semay, C. Roland, Approximate solutions for $N$-body Hamiltonians with identical particles in $D$ dimensions, Res. Phys. \textbf{3}, 231 (2013)
\bibitem{chev21} C. Chevalier, C.T. Willemyns, L. Cimino, C. Semay, Improvement of the Envelope Theory for Systems with Different Particles, Few-Body Syst \textbf{63}, 40 (2022)
\bibitem{schr15} M. Schr\"oter \emph{et al.}, Exciton-vibrational coupling in the dynamics and spectroscopy of Frenkel excitons in molecular aggregates, Phys. Rep. \textbf{567}, 1 (2015)
\bibitem{mara18} A. Marais \emph{et al.}, The future of quantum biology, J. R. Soc. Interface \textbf{15}, 20180640 (2018)
\bibitem{gras17} F. Grasselli, Variational approach to the soft-Coulomb potential in low-dimensional quantum systems, Am. J. Phys. \textbf{85}, 834 (2017)
\bibitem{sema19} C. Semay, L. Cimino, Tests of the Envelope Theory in One Dimension, Few-Body Syst \textbf{60}, 64 (2019)
\bibitem{cimi21} L. Cimino, C. Semay, Compact Equations for the Envelope Theory, Braz. J. Phys. \textbf{52}, 45 (2022)
\bibitem{sema16} C. Semay, L. Ducobu, Quantum and classical probability distributions for arbitrary Hamiltonians, Eur. J. Phys. \textbf{37}, 045403 (2016)
\bibitem{sema15a} C. Semay, Numerical Tests of the Envelope Theory for Few-Boson Systems, Few-Body Syst \textbf{56}, 149 (2015)
\bibitem{sema11} C. Semay, General comparison theorem for eigenvalues of a certain class of Hamiltonians, Phys. Rev. A \textbf{83}, 024101 (2011)
\bibitem{jara09} B. Jaramillo, R.P. Mart\'inez-y-Romero, H.N. N\'u\~nez-Y\'epez, A.L. Salas-Brito, On the one-dimensional Coulomb problem, Phys. Lett. A \textbf{374}, 150 (2009)
\bibitem{mars89} C. Clay Marston, G.G. Balint-Kurti, The Fourier grid Hamiltonian method for bound state eigenvalues and eigenfunctions, J. Chem. Phys. \textbf{91}, 3571 (1989)
\bibitem{sema00} C. Semay, Fourier grid Hamiltonian method and Lagrange-mesh calculations, Phys. Rev. E \textbf{62}, 8777 (2000)
\bibitem{macd33} J.K.L. MacDonald, Successive Approximations by the Rayleigh-Ritz Variation Method, Phys. Rev. \textbf{43}, 830 (1933)
\bibitem{buis12} F. Buisseret, N. Matagne, C. Semay, Spin contribution to light baryons in different large-$N$ limits, Phys. Rev. D \textbf{85}, 036010 (2012)
\bibitem{sema21} C. Semay, C.T. Willemyns, Quasi Kepler's third law for quantum many-body systems, Eur. Phys. J. Plus \textbf{136}, 342 (2021)
\bibitem{buis22} F. Buisseret, C.T. Willemyns, C. Semay, Many-Quark Interactions: Large-$N$ Scaling and Contribution to Baryon Masses, Universe \textbf{8}, 311 (2022)
\bibitem{horn14} J. Horne, J.A. Salas, K. Varga, Energy and Structure of Few-Body Systems, Few-Body Syst \textbf{55}, 1245 (2014)
\bibitem{timo17} N.K. Timofeyuk, D. Baye, Hyperspherical Harmonics Expansion on Lagrange Meshes for Bosonic Systems in One Dimension, Few-Body Syst \textbf{58}, 157 (2017)
\bibitem{sema15b} C. Semay, Improvement of the envelope theory with the dominantly orbital state method, Eur. Phys. J. Plus \textbf{130}, 156 (2015)
\bibitem{delv21} J.C. del Valle, A.V. Turbiner, Power-like potentials: From the Bohr Sommerfeld energies to exact ones, Int. J. Mod. Phys. A \textbf{36}, 2150221 (2021) 
\bibitem{sing85} D. Singh, Y.P. Varshni, R. Dutt, Bound eigenstates for two truncated Coulomb potentials, Phys. Rev. A \textbf{32}, 619 (1985)

\end{thebibliography}
\end{document}